\documentclass[twocolumn,showpacs,preprintnumbers,amsmath,amssymb]{revtex4}
\usepackage{graphicx}% Include figure files
\usepackage{dcolumn}% Align table columns on decimal point
\usepackage{bm}% bold math
\usepackage{amsthm}
\begin{document}
\title{\bf{Local Entanglement Is Not Necessary for Perfect Discrimination between  Unitary Operations Acting on Two-Qudits by
 LOCC}}
\author{Lvzhou Li}\email{lilvzhou@mail2.sysu.edu.cn
(L. Li).}  \author{Daowen Qiu}\email{issqdw@mail.sysu.edu.cn (D.
Qiu).}%

 \affiliation{%
 Department of
Computer Science, Zhongshan University, Guangzhou 510275,
 People's Republic of China
}%

\date{\today}

 \begin{abstract}
Recently, the problem of discriminating  multipartite unitary
operations by local operations and classical communication (LOCC)
has attracted significant attention. The latest work in the
literature on this problem showed that two multipartite unitary
operations can always be perfectly distinguished by LOCC when a
finite number of runs are allowable. However, in these schemes,
local entanglement (an entangled state holden by one party) was
required, which seems to imply that local entanglement is
necessary for perfect discrimination between unitary operations by
LOCC.  In this article, we show that a perfect discrimination
between two unitary operations acting on a two-qudits can always
be achieved without exploiting any entanglement. As a result, we
conclude that local entanglement is not necessary for perfect
discrimination between unitary operations acting on two-qudits by
LOCC.
 \end{abstract}
  \maketitle
{\it Introduction}---
 The quantum nonorthogonality and entanglement are at the heart of
 quantum information. The former has
 close relation with the distinguishability of quantum states, which
 has been extensively studied by \cite{Hel,Four} and others, and a
 recent review on this is referred to \cite{Bergou04}. On the other hand,
 quantum entanglement, playing a fundamental role in quantum
 computation and information, is closely relative to the quantum
 nonlocality. Recently, Bennett {\it et al}  \cite{Ben99} found a
 surprising
 phenomenon--- ``quantum nonlocality without entanglement'', which
 exhibits  a set of orthogonal bipartite pure product states that
 can not be perfectly distinguished by only local operations and
 classical communication (LOCC). Then, inspired by the seminal idea \cite{Ben99}, many works have been devoted
 to the link between  quantum distinguishability and  quantum
 nonlocality, and some related problems. Specially,
 Walgate {\it et al}  \cite{Walgate} showed that any two
 orthogonal entangled states can be perfectly distinguished by
 LOCC, which implies
 that some nonlocality can be recovered by only local operations.

The similar problems can also be considered for quantum
operations. Thus, the discrimination of quantum operations has
attracted many authors (for example, \cite{MFS2,
Chef07,Childs,Acin,Paris, Duan1,Duan2, Zhou}).
 In this article, we focus on  the discrimination of unitary
operations  \cite{Childs,Acin,Paris, Duan1,Duan2, Zhou}. Two
unitary operations $U$ and $V$ are said to be perfectly
distinguishable, if there exists an input state $|\psi\rangle$
such that $U|\psi\rangle\perp V|\psi\rangle$. The already known
works on this problem can be divided into two lines, and we will
briefly recall them below.

The first line is the works by \cite{Childs,Acin,Paris, Duan1},
where the  unitary operations to be discriminated are under the
complete control of a single party who can  perform any physically
allowed operations to achieve an optimal discrimination. Firstly,
Refs.~\cite{Acin,Paris} showed that two unitary operations $U$ and
$V$ can be perfectly discriminated with only single run allowed
if, and only if $\Theta(U^\dagger V)\geq\pi$, where $\Theta(U)$
denotes the length of the smallest arc containing all the
eigenvalues of $U$ on the unit circle. However,
 the situation changes dramatically when a finite number of runs  are allowed. Specifically,
Refs.~\cite{Acin,Paris} showed that for any two different unitary
operations $U$ and $V$, there exists a finite number $N$ such that
$U^{\otimes N}$ and $V^{\otimes N}$ can be perfectly
discriminated. Intuitively, such a discriminating scheme  is
called a {\it parallel scheme}. It is worth pointing out that in
the parallel scheme, an $N$-partite entangled state is necessary
and plays a key role. Latterly, this result was further refined
in~\cite{Duan1} by showing that the entangled input is not
necessary. Specially, ~\cite{Duan1} showed that for any two
different unitary operations $U$ and $V$, there exist input state
$|\varphi\rangle$ and auxiliary operations $X_1,\dots,X_N$ such
that $UX_NU\dots X_1U|\varphi\rangle\perp VX_NV\dots
X_1V|\varphi\rangle$. Generally speaking, we call such a
discriminating scheme as a {\it sequential scheme}.

The second line is the recent works  \cite{Duan2,Zhou}, where the
 unitary operations to be discriminated   are shared by
several spatially separated parties. Thus, a reasonable constraint
on the discrimination is that each party can only make local
operations and classical communication (LOCC). Clearly, the
problem becomes more complicated in this case. Amazedly,
Refs.~\cite{Duan2, Zhou} independently showed that if  a finite
number  of runs are allowed, then any two different unitary
operations can be perfectly discriminated by LOCC.
Refs.~\cite{Duan2,Zhou} used different methods to achieve the same
finding. For instance,
  \cite{Duan2} was mainly based on the analysis of
numerical range \cite{Horn}, while  \cite{Zhou} mainly made use of
the result on the university of quantum gate \cite{Bry}. Although
the methods used in \cite{Duan2, Zhou} are different, the main
idea of them is similar and can be summarily described  as
follows:

 (i) For two bipartite unitary operations $U$ and $V$ shared
by Alice and Bob, which satisfy some special condition, it is
showed that there exists a finite number $N$ such that $U^{\otimes
N}$ and $V^{\otimes N}$ can be discriminated by such a product
state $|\varphi\rangle_A|\psi\rangle_B$ where $|\varphi\rangle_A$
and $|\psi\rangle_B$ are two N-partite states holden by Alice and
Bob, respectively, and one of which must be an N-partite entangled
state. In this article, we call such an entangled state holden by
one party as {\it local entanglement}.

 (ii) For any two general bipartite unitary
operation $U$ and $V$ to be discriminated, we construct two
quantum circuits $f(X)=Xw_1X\dots w_nX$ with $X\in\{U,V\}$ by
finding a suitable sequence of local unitary operations
$w_1,\dots,w_n$ where each $w_i$ has this form $w_i=u_i\otimes
v_i$, such that $f(U)$ and $f(V)$ satisfy the desired condition
stated in step (i). Thus $f(U)$ and $f(V)$ can be discriminated as
in step (i), which means that $U$ and $V$ can be perfectly
discriminated by LOCC.

\begin{figure}
 \setlength{\unitlength}{1.7mm}
  \begin{picture}(50,23)

\put(-1,15.5){Alice}
 \put(8,20){\line(-1,-1){4}}
\put(8,6){\line(-2,5){4}}

\put(-1,5.5){Bob}
 \put(8,16){\line(-2,-5){4}}
\put(8,2){\line(-1,1){4}}

 \put(8,20){\line(1,0){2}}
 \put(8,16){\line(1,0){2}}
\put(10,15){\framebox(4,6){$X$}}

 \put(14,20){\line(1,0){2}}
 \put(14,16){\line(1,0){2}}
\put(16,18.5){\framebox(2.5,2.5){$u_1$}}
\put(16,14.5){\framebox(2.5,2.5){$v_1$}}
\put(18.5,20){\line(1,0){2}}
 \put(18.5,16){\line(1,0){2}}
 \put(20.5,15){\framebox(4,6){$X$}}
 \put(24.5,20){\line(1,0){1}}
 \put(24.5,16){\line(1,0){1}}
\put(26,20){\line(1,0){0.5}}
 \put(26,16){\line(1,0){0.5}}
 \put(27,20){\line(1,0){0.5}}
 \put(27,16){\line(1,0){0.5}}
\put(28,20){\line(1,0){0.5}}
 \put(28,16){\line(1,0){0.5}}
 \put(29,20){\line(1,0){1}}
 \put(29,16){\line(1,0){1}}
 \put(30,18.5){\framebox(2.5,2.5){$u_n$}}
\put(30,14.5){\framebox(2.5,2.5){$v_n$}}
\put(32.5,20){\line(1,0){2}}
 \put(32.5,16){\line(1,0){2}}
 \put(34.5,15){\framebox(4,6){$X$}}
 \put(38.5,20){\line(1,0){2}}
 \put(38.5,16){\line(1,0){2}}

  \put(22.5,13){\circle*{0.4}}
  \put(22.5,11){\circle*{0.4}}
  \put(22.5,9){\circle*{0.4}}
 \put(42,11){$|\Phi_X\rangle$}
  \put(8,6){\line(1,0){2}}
 \put(8,2){\line(1,0){2}}
\put(10,1){\framebox(4,6){$X$}}

 \put(14,6){\line(1,0){2}}
 \put(14,2){\line(1,0){2}}
\put(16,4.5){\framebox(2.5,2.5){$u_1$}}
\put(16,0.5){\framebox(2.5,2.5){$v_1$}}
\put(18.5,6){\line(1,0){2}}
 \put(18.5,2){\line(1,0){2}}
 \put(20.5,1){\framebox(4,6){$X$}}
 \put(24.5,6){\line(1,0){1}}
 \put(24.5,2){\line(1,0){1}}
\put(26,6){\line(1,0){0.5}}
 \put(26,2){\line(1,0){0.5}}
 \put(27,6){\line(1,0){0.5}}
 \put(27,2){\line(1,0){0.5}}
\put(28,6){\line(1,0){0.5}}
 \put(28,2){\line(1,0){0.5}}
 \put(29,6){\line(1,0){1}}
 \put(29,2){\line(1,0){1}}
 \put(30,4.5){\framebox(2.5,2.5){$u_n$}}
\put(30,0.5){\framebox(2.5,2.5){$v_n$}}
\put(32.5,6){\line(1,0){2}}
 \put(32.5,2){\line(1,0){2}}
 \put(34.5,1){\framebox(4,6){$X$}}
 \put(38.5,6){\line(1,0){2}}
 \put(38.5,2){\line(1,0){2}}
  \end{picture}
\caption   { \label{Fig 2}A mixed scheme to distinguish bipartite
unitary operations $X\in\{U,V\}$ by LOCC. For a perfect
discrimination, one of Alice and Bob must prepare an entangled
input state, i.e.,  local entanglement.}
\end{figure}
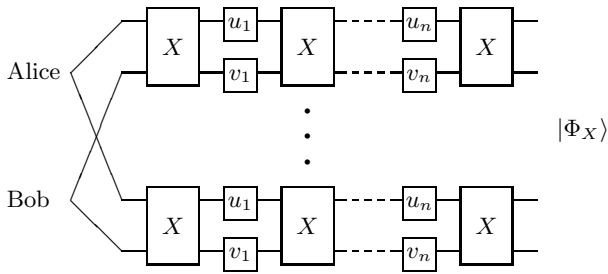

Also, we can  use Fig.~\ref{Fig 2} to visualize the above idea.
Then as we can see,  we should  generally  combine the parallel
scheme and the sequential scheme stated before  to achieve a
perfect discrimination between two bipartite unitary operations by
LOCC. Intuitively, we call such a process as a {\it mixed scheme}.
Again, it is worth  pointing out that  in the above process, one
party must prepare  local entanglement that is essentially an
entangled state shared by several subsystems holden by one party,
which seems to imply that {\it local entanglement is necessary for
perfect discrimination between unitary operations by LOCC}.

Naturally,  there is   a  question to be addressed: {\it is the
local entanglement  indubitably necessary ?} We think that this
question is  nontrivial   in the sense of both practice and theory
as follows. (a) Local entanglement is essentially an entangled
state shared by several subsystems, which  presents nonlocality
among  these subsystems, and, as a valuable physical resource, is
generally difficult to prepare. Consequently,  in practice it is
of great importance to accomplish a given task without exploiting
any entanglement as possible as we can. (b) Theoretically
speaking, it is greatly significative to find out what kind of
tasks can be achieved without exploiting any entanglement, since
it was still argued that it may be the interference and the
orthogonality but not the entanglement which are responsible for
the power of quantum computing \cite{Meyer}. Besides the above
question, another natural question is
 whether there exists a simpler protocol using merely the
parallel scheme or the sequential scheme to achieve the perfect
discrimination of two unitary operations by LOCC.

In this article, we will answer the two questions by showing that
any two bipartite unitary operations acting on a $d\otimes d$
Hilbert space (i.e., two qudits; a qudit is a $d$-dimensional
quantum system), allowed with a finite number of runs, in
principle, can be locally distinguished with certainty by a
sequential scheme without  exploiting any entanglement. Then, we
will obtain this statement: {\it local entanglement is not
necessary for perfect discrimination between unitary operations
acting on two-qudits by LOCC}, which is a stronger result than
that in ~\cite{Duan1}---``entanglement is not necessary for
perfect discrimination between unitary operations''. Consequently,
this will  be a new instance of the kind of tasks which can be
achieved without employing entanglement \cite{Duan1}.

{\it Preliminaries}--- Here some useful results and notation are
introduced. Since the problem of discriminating unitary operations
by LOCC is generally transformed to the problem of discriminating
quantum states by LOCC, we first recall a fundamental result by
Walgate {\it et al} \cite{Walgate} as follows.

{\it Lemma 1.} Let $|\varphi\rangle_1$ and  $|\varphi\rangle_2$ be
two orthogonal multipartite  pure states. Then $|\varphi\rangle_1$
and $|\varphi\rangle_2$ are perfectly distinguishable by LOCC.

Note that we say unitary operations $U$ and $V$ are different if
$U\neq
 e^{i\theta}
V$ for any real $\theta$, and for simplicity, we always denote
that by $U\neq V$.  Next we recall another useful result regarding
the distinguishability of unitary operations in \cite{Duan1}.

{\it Lemma 2.} Let $U$ and $V$ be two different unitary
operations, and let $N=\lceil\frac{\pi}{\Theta(U^\dagger
V)}\rceil-1$. Then there exist auxiliary unitary operations
$X_1,\dots,X_{N}$ and input state $|\psi\rangle$ such that
\begin{align}
UX_{N}U\dots X_1U|\psi\rangle\perp VX_{N}V\dots
X_1V|\psi\rangle.\label{sequential}
\end{align}

The above scheme is the so-called sequential scheme for
discriminating two unitary operations.  Now let us have a further
analysis  on this scheme. Suppose that the two operations $U_{AB}$
and $V_{AB}$ to be discriminated are unitary operations acting on
two qudits. Then, in terms of the proof of \cite{Duan1}, we can
see that the auxiliary operations $X_i$ are generally global
operations acting on the two qudits, and the input  state
$|\varphi\rangle$ is generally an entangled state of the two
qudits. Therefore, intuitively, this scheme will be not valid for
discriminating two bipartite operations $U_{AB}$ and $V_{AB}$ if
the two qudits are spatially separated, since then only local
operations and classical communication are feasible. However, we
may ask this question: for the bipartite operations $U_{AB}$ and
$V_{AB}$, do there exist some local unitary operations in the form
$X_i=X_i^A\otimes X_i^B$ and a product input state
$|\varphi\rangle=|\alpha\rangle_A|\beta\rangle_B$ such that
Eq.~\eqref{sequential} holds?  Indeed, we can prove that such a
scheme does exist, and thus, we can also address the two questions
raised in the Introduction.

We now focus on the distinguishability of multipartite unitary
operations by LOCC. For simplicity, we consider unitary operations
acting on a two-qudits  (as mentioned before, a qudit is a
$d$-dimensional quantum system). Let ${\cal H}_d$ denote the state
space of a qudit system. Then the state space of a 2-qudit system
is denoted by ${\cal H}={\cal H}_d\otimes{\cal H}_d$. Sometimes,
we will use $d\otimes d$ as an abbreviation  for ${\cal H}$. The
sets of unitary operations acting on two qudits and on a single
qudit are denoted by  ${\cal U}({\cal H})$, and ${\cal U}({\cal
H}_d)$, respectively.

According to \cite{Bry}, we call $U\in{\cal U}({\cal H})$ to be
{\it primitive} if it maps a separable state to another separable
state, i.e., for any qudit states $|x\rangle$ and $|y\rangle$, we
can find  qudit states $|u\rangle$ and $|v\rangle$ such that
$U|x\rangle|y\rangle=|u\rangle|v\rangle$. Otherwise, it is {\it
imprimitive}. For the primitive operations, we have this
characterization \cite{Bry}: $U\in{\cal U}({\cal H})$ is primitive
if and only if $U=U_A\otimes U_B$ or $U=(U_A\otimes U_B)P$, where
$P$ is a swap operation, i.e.,
$P|x\rangle|y\rangle=|y\rangle|x\rangle$. For simplicity, we use
$S$ to denote the set of all 2-qudit unitary operations in the
form $U_A\otimes U_B$. With these notation, we introduce the
following lemma.

{\it Lemma 3.} $S$ together with an imprimitive operation $Q$ can
generate any  unitary operation acting on a two-qudit system.

This lemma was proven in detail by \cite{Bry}. Specifically, for
an imprimitive operation $Q$, by constructing $S^{'}=QSQ^{-1}$,
and then by choosing a suitable sequence of $S$ and $S^{'}$, we
can obtain any desired element in ${\cal U}({\cal H})$. Note that
the sequence of $S$ and $S^{'}$ generally has this form
$(SS^{'})^nS$.

{\it The main result}---
 Now, we are  in a position to deal with
the problem of discriminating unitary operations by LOCC, which is
formalized as follows:

  {\bf Problem:} {\it For any two different  operations $U,
V\in{\cal U}({\cal H})$ shared by Alice and Bob and allowed with a
finite number of runs, can we find a product input state
$|\varphi\rangle_A|\psi\rangle_B\in{\cal H}$, and  quantum
circuits $f(X)$  built upon some local operations and
$X\in\{U,V\}$, such that $f(U)|\varphi\rangle_A|\psi\rangle_B\perp
f(V)|\varphi\rangle_A|\psi\rangle_B$}?

 Before dealing with the problem, a useful observation should be pointed out.
 In our problem,
a unitary operation $U$ can   be regarded as a black box with some
input and output ports, irrespective of its inner complexity, and
then, by exchanging the input and output ports of the whole setup,
we can soon obtain the reverse transformation $U^\dagger$. Thus,
as long as unitary operation $U$ is available, $U^\dagger$ is also
available, and using $U^\dagger$ can be taken as using $U$. This
fact will be always  used in the following discussion, and it will
be useful in our proof.

Now, we give our main result in the following.

{\it Theorem 1.} Any two different unitary operations acting on a
2-qudit system, allowed with a finite number of runs, can always
be locally  perfectly discriminated by a sequential scheme without
exploiting any entanglement.

 {\it Proof.} We prove this theorem by dealing with the above
 problem in the following three cases:

Case (i): $U$ and $V$ are all primitive. Then it  suffices to
consider the  following three subcases:

Case (i-a): $U=U_A\otimes U_B$ and $V=V_A\otimes V_B$. Without
lose of generality, assume that $U_A\neq  V_A$. Then we can simply
discriminate $U$ and $V$ by discriminating $U_A$ and $V_A$. From
Lemma 2, it is easy to see that there exist $X_1,\dots,
X_N\in{\cal U}({\cal H}_d)$ and $|\psi\rangle_A\in {\cal H}_d$
such that
\begin{align*}
&U(X_N\otimes I)U\dots (X_1\otimes
I)U|\psi\rangle_A|\varphi\rangle_B \perp\\ &V(X_N\otimes I)V\dots
(X_1\otimes I)V|\psi\rangle_A|\varphi\rangle_B.
\end{align*}
 Therefore, $U$ and $V$ can be
discriminated by LOCC.

Case (i-b): $U=U_A\otimes U_B$ and $V=(V_A\otimes V_B)P$. Let
$|\varphi\rangle|\phi\rangle\in{\cal H}$. Then we have
\begin{align*}
|r\rangle=U|\varphi\rangle|\phi\rangle=U_A|\varphi\rangle\otimes
U_B|\phi\rangle,\\
|r^{'}\rangle=V|\varphi\rangle|\phi\rangle=V_A|\phi\rangle\otimes
V_B|\varphi\rangle,
\end{align*}
and
\begin{align*}
\langle r|r^{'}\rangle=\langle\varphi|U_A^\dagger
V_A|\phi\rangle\langle\phi|U_B^\dagger U_B|\varphi\rangle.
\end{align*}
It is readily seen that we can let $|\phi\rangle=V_A^\dagger
U_A|\varphi^\perp\rangle$ where $|\varphi^\perp\rangle$ denotes a
state orthogonal to $|\varphi\rangle$, such that $\langle
r|r^{'}\rangle=0$. Therefore, $U$ and $V$ can be discriminated by
LOCC.

Case (i-c): $U=(U_A\otimes U_B)P$ and $V=(V_A\otimes V_B)P$.
Without loss of generality, assume that $U_A\neq V_A$. Let
\begin{align*}
f(U)=U(X_1\otimes X_2)U^\dagger, \text{~and~} f(V)=V(X_1\otimes
X_2)V^\dagger,
\end{align*}
 where $X_1$ and $X_2$ are two fixed elements in
${\cal U}({\cal H}_d)$. Then for any product state
$|\varphi\rangle|\phi\rangle\in{\cal H}$, by straight calculation,
we have
\begin{align*}
f(U)|\varphi\rangle|\phi\rangle=U_AX_2U_A^\dagger\otimes U_BX_1U_B^\dagger|\varphi\rangle|\phi\rangle,\\
f(V)|\varphi\rangle|\phi\rangle=V_AX_2V_A^\dagger\otimes
V_BX_1V_B^\dagger|\varphi\rangle|\phi\rangle.
\end{align*}
Thus, by the linearity of unitary operations, we obtain that:
\begin{align*}
f(U)=U_AX_2U_A^\dagger\otimes U_BX_1U_B^\dagger,\\
f(V)=V_AX_2V_A^\dagger\otimes V_BX_1V_B^\dagger.
\end{align*}
Since $U_A\neq V_A$, we have $U^\dagger_AV_A\neq I$ (up to any
phase factor). Thus there exists suitable $X_2$ such that
$X_2U^\dagger_AV_A\neq U^\dagger_AV_A X_2$, i.e.,
$U_AX_2U_A^\dagger\neq V_AX_2V_A^\dagger$. Therefore, by the
discussion in subcase (i-a), we can discriminate $f(U)$ and $f(V)$
and thus discriminate $U$ and $V$ by LOCC.

Case (ii): One of $U$ and $V$ is primitive. Without loss of
generality, assume that $V$ is primitive. Then we discuss that by
the following two subcases:

Case (ii-a): $V=V_A\otimes V_B$ and $U$ is imprimitive.  In terms
of Lemma 3, we can construct a quantum circuit
$f(X)\in(SS^{'})^nS$ where $S^{'}=XSX^{-1}$, such that
\begin{align*}f(U)=P_A\otimes I_B+P^{'}_A\otimes
U_B^{'},\end{align*}  where $U_B^{'}\neq I_B$, and $P_A$ and
$P^{'}_A$ are two projectors and $P_A\oplus P_A^{'}=I_A$. In other
words, $f(U)$ is a controlled unitary transformation. At the same
time, it is clear that $f(V)\in S$. Thus we let
$f(V)=V_A^{'}\otimes V_B^{'}$, where we should have either
$V^{'}_B\neq I_B$ or $V^{'}_B\neq U^{'}_B$. Without loss of
generality, assume that $V_B^{'}\neq U_B^{'}$. In the following,
we can see that $f(U)$ and $f(V)$ can be discriminated by LOCC.
Primarily, we have a lemma as follows.

{\it Lemma 4.} For the controlled unitary transformation
$U=P_A\otimes I_B+P^{'}_A\otimes U_B^{'}$, let
$|\alpha\rangle_A\in{\cal H}_d$ satisfy
$P_A^{'}|\alpha\rangle_A=|\alpha\rangle_A$. Then for any
$|\varphi\rangle_B\in{\cal H}_d$, and $X_1,\dots, X_N\in{\cal
U}({\cal H}_d)$, we have
\begin{align*}
&U(I\otimes X_N)U\dots(I\otimes
X_1)U|\alpha\rangle_A|\varphi\rangle_B\\=&|\alpha\rangle_A\otimes(U_B^{'}X_NU_B^{'}\dots
X_1U_B^{'})|\varphi\rangle_B.
\end{align*}

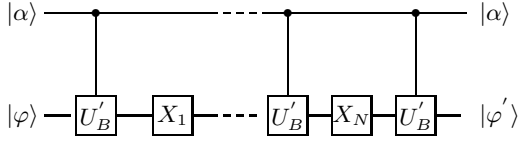
\begin{figure}
 \setlength{\unitlength}{1.7mm}
  \begin{picture}(50,18)

 \put(2,12){$|\alpha\rangle$}
 \put(2,4){$|\varphi\rangle$}

 \put(5,12.5){\line(1,0){13.5}}
 \put(19,12.5){\line(1,0){0.5}} \put(20,12.5){\line(1,0){0.5}}
\put(21,12.5){\line(1,0){0.5}}
 \put(9,12.5){\circle*{0.7}}
 \put(9,12.5){\line(0,-1){6.5}}
  \put(7.5,3){\framebox(3,3){$U_B^{'}$}}
 \put(5,4.5){\line(1,0){2}}
 \put(10.5,4.5){\line(1,0){3}}
\put(13.5,3){\framebox(3,3){$X_1$}}
 \put(16.5,4.5){\line(1,0){2}}
\put(19,4.5){\line(1,0){0.5}} \put(20,4.5){\line(1,0){0.5}}
\put(21,4.5){\line(1,0){0.5}}

\put(22,12.5){\line(1,0){16}} \put(24,12.5){\circle*{0.7}}
 \put(24,12.5){\line(0,-1){6.5}}
 \put(22.5,3){\framebox(3,3){$U_B^{'}$}}
  \put(25.5,4.5){\line(1,0){2}}
 \put(27.5,3){\framebox(3,3){$X_N$}}
  \put(30.5,4.5){\line(1,0){2}}
   \put(32.5,3){\framebox(3,3){$U_B^{'}$}}
   \put(35.5,4.5){\line(1,0){2}}
   \put(34,12.5){\circle*{0.7}}
 \put(34,12.5){\line(0,-1){6.5}}
\put(39,12){$|\alpha\rangle$} \put(39,4){$|\varphi^{'}\rangle$}
  \end{picture}
\caption   { \label{Fig 1}For the controlled unitary
transformation $U=P_A\otimes I_B+P^{'}_A\otimes U_B^{'}$,
construct such a quantum circuit. Then, by inputting
$|\alpha\rangle$ satisfying $P_A^{'}|\alpha\rangle=|\alpha\rangle$
 to the control qudit, and $|\varphi\rangle$ to the target qudit,
 we get the control qudit invariable and the target qudit being
 $|\varphi^{'}\rangle=U_B^{'}X_NU_B^{'}\dots
X_1U_B^{'})|\varphi\rangle$.}
\end{figure}

 The proof  of  this lemma is easy, and we visualize it in Fig \ref{Fig 1}. With this lemma, we can now discriminate $f(U)$ and $f(V)$ by
discriminating $U_B^{'}$ and $V_B^{'}$ as follows. According to
Lemma 2,  there exist $|\varphi\rangle_B\in{\cal H}_d$, and
$X_1,\dots, X_N\in{\cal U}({\cal H}_d)$ such that
$$U_B^{'}X_NU_B^{'}\dots X_1U_B^{'}|\varphi\rangle_B\perp V_B^{'}X_NV_B^{'}\dots X_1V_B^{'}|\varphi\rangle_B.$$
Thus, by Lemma 4, we have
\begin{align*}
&f(U)(I\otimes X_N)f(U)\dots(I\otimes
X_1)f(U)|\alpha\rangle_A|\varphi\rangle_B\perp\\&f(V)(I\otimes
X_N)f(V)\dots(I\otimes X_1)f(V)|\alpha\rangle_A|\varphi\rangle_B.
\end{align*}
Therefore, $f(U)$ and $f(V)$ can be  discriminated by LOCC, i.e.,
$U$ and $V$ can be  discriminated by LOCC.

Case (ii-b): $V=(V_A\otimes V_B)P$ and $U$ is imprimitive.  This
is similar to the subcase (ii-a) by noting that $VSV^{-1}= S$, and
thus $f(V)\in S$.

Case (iii): Both $U$ and $V$ are  imprimitive. The proof for this
case is some more complicated than before, and we need a useful
lemma in the following.

{\it Lemma 5.} For  unitary operation $U$ acting on  $d\otimes d$,
$U^\dagger=AUA^\dagger$ holds for any $A\in {\cal
U}=\{(\sigma_z\oplus I)\otimes I, (\sigma_y\oplus I)\otimes
I,I\otimes(\sigma_z\oplus I),I\otimes(\sigma_y\oplus I)\}$ if, and
only if $U$ has this form $U=e^{ixu_1\otimes u_2}$ for some real
number $x$, where $u_1=u_2=\sigma_x\oplus 0_{(d-2)}$ and
$\sigma_x$, $\sigma_y$ and $\sigma_z$ are Pauli operators.

This result was also used by \cite{Zhou}, however, without giving
a rigorous  proof.  In view of its nontrivial role in \cite{Zhou}
and this article, we will give a detailed proof for it. But for
the continuity of the proof for Theorem 1,  we just accept this
result at the moment, and we will present its proof in a separate
paragraph subsequently.

With Lemma 5, we can now prove Case (iii) as follows. Firstly, by
Lemma 3 we can construct $f(X)\in(SS^{'})^nS$ where
$S^{'}=XSX^{-1}$, such that $f(U)=e^{iu_1\otimes u_2}$ with
$u_1=u_2=\sigma_x\oplus {\bf 0}$. It is easy to check that $f(U)$
is imprimitive. Now, if $f(V)$ is primitive, then by the
discussion in Case (ii), we know that $f(U)$ and $f(V)$ can be
discriminated by LOCC. Otherwise, based on Lemma 5, we have the
following considerations:

Case (iii-a):  $f(V)\neq e^{ixu_1\otimes u_2}$.  Let
$F(X)=Af(X)A^\dagger f(X)$ for some $A\in{\cal U}$. Then by Lemma
5, we have $F(U)=I$ and $F(V)\neq I$ for some $A$. Therefore, by
the previous discussion, $F(U)$ and $F(V)$ can be discriminated by
LOCC.

Case (iii-b): $f(V)= e^{ixu_1\otimes u_2}$. When $x=1$, $f(U)$ and
$f(V)$ are the same and imprimitive.  Thus by Lemma 3, we can
construct a quantum circuit $h(.)$ such that $h(f(U))=U^\dagger$,
and then we have $Uh(f(U))=I$ and $Vh(f(V))=VU^\dagger$.
Therefore, they can be discriminated by LOCC from the previous
discussion. When $x\neq 1$,  discriminating $f(U)$ and $f(V)$ can
be reduced to discriminating $e^{iu_1}$ and $e^{ixu_1}$ as
follows. By inputting $|\varphi\rangle_A|\alpha\rangle_B$ where
$|\alpha\rangle_B$ is an eigenvector of $u_2$ corresponding with
eigenvalue $1$, it is easy to check that $e^{ixu_1\otimes
u_2}|\varphi\rangle_A|\alpha\rangle_B=(e^{ixu_1}\otimes
I)|\varphi\rangle_A|\alpha\rangle_B$. Furthermore, we have
\begin{align*}
|r\rangle&=f(U)(X_N\otimes I)f(U)\dots(X_1\otimes
I)f(U)|\varphi\rangle_A|\alpha\rangle_B\\&=(e^{iu_1}X_Ne^{iu_1}\dots
X_1e^{iu_1})|\varphi\rangle_A\otimes|\alpha\rangle_B,\\
 |r^{'}\rangle&=f(V)(X_N\otimes I)f(V)\dots(X_1\otimes
I)f(V)|\varphi\rangle_A|\alpha\rangle_B\\&=(e^{ixu_1}X_Ne^{ixu_1}\dots
X_1e^{ixu_1})|\varphi\rangle_A\otimes|\alpha\rangle_B.
\end{align*}
Therefore, with Lemma 2, by choosing suitable input state
$|\varphi\rangle_A$ and auxiliary operations $X_i$, we can make
$|r\rangle\perp|r^{'}\rangle$. Thus, $f(U)$ and $f(V)$ can be
discriminated by LOCC.

From the above discussion, one can see that our basic idea in the
proof can be summarized as follows:  (i)  construct a sequential
circuit by Lemma 2 or Lemma 3; (ii)  embed this circuit in another
sequential circuit; (iii) repeat the above two steps finite times,
obtaining the final circuit which  is clearly sequential and  can
be used to discriminate $U$ and $V$ by LOCC.  Also, it is easy to
see that the above process does not employ any entanglement, based
on the following two points: (a) the input state is a bipartite
product state which does not present any entanglement; (b) one can
find that in each case of the above proof, the  output states
corresponding to the two unitary operations to be discriminated,
are two orthogonal {\it product} states which can be easily
discriminated by local operations without  involving any auxiliary
system \cite{Rm}. Thus, we have completed the proof of Theorem
1.\qed

So far, we have considered the problem of discriminating two
bipartite unitary operations acting on a two-qudits by LOCC. In
particular, we obtain this result: local entanglement is not
necessary for perfect discrimination between unitary operations
acting on a two-qudits by LOCC. Now, we can generalize this result
to the case of $N$ unitary operations acting on two-qudits, by
performing the above discriminating process $N-1$ times as did in
\cite{Acin,Duan1}. Besides, there are some open problems worthy of
further consideration. The first problem is how to deal with the
case that the two subsystems $A$ and $B$ have different
dimensions. (In Theorem 1, we have considered only the case that
the two subsystems have equal dimension.) Another challenging
problem is how to extend our result to the case of multipartite
unitary operations.

Last but not least, it is the complexity of a discriminating
scheme that we should consider. In the global scene
\cite{Acin,Paris, Duan1}, it has  been shown that
$N=\lceil\frac{\pi}{\Theta(U^\dagger V)}\rceil$ is the optimal
number of runs for a perfect discrimination between $U$ and $V$,
which implies that in the LOCC scene, the optimal number of runs
$N^{'}$ satisfies $N^{'}\geq N$. In some special cases, we may
have $N^{'}=N$; for instance, in Case (i-b) of our discussion, we
can get that $N^{'}=N$, since there is $N^{'}=1$. However, what is
the sufficient and necessary condition for $N^{'}=N$ is still left
open. Furthermore, what is the exact expression of $N^{'}$  is
still unknown. We hope these problems will be addressed in the
further study.

{\it The proof of Lemma 5}---Here we will give a detailed proof
for Lemma 5. Firstly, we have the following fact.

{\it Fact 1.}
 Any  unitary operation $U$ can be expressed as $U=e^{iB}$ where
 $B $ is Hermitian, and for unitary operation $A$,
 $AUA^\dagger=U^\dagger$ if and only if
 $B=-ABA^\dagger$.

 This fact can be easily proven, and it will be useful later. In
 the interest of
 readability,  let us recall  the statement of Lemma 5 before beginning its proof.

 {\it Statement of Lemma 5}: For unitary operation $U$ acting on  $d\otimes
d$, $U^\dagger=AUA^\dagger$ holds for any $A\in {\cal
U}=\{(\sigma_z\oplus I)\otimes I, (\sigma_y\oplus I)\otimes
I,I\otimes(\sigma_z\oplus I),I\otimes(\sigma_y\oplus I)\}$ if, and
only if $U$ has this form $U=e^{ixu_1\otimes u_2}$ for some real
number $x$, where $u_1=u_2=\sigma_x\oplus 0_{(d-2)}$ and
$\sigma_x$, $\sigma_y$ and $\sigma_z$ are Pauli operators.

{\it Proof of Lemma 5:}
 Firstly, we verify the sufficiency. Suppose that
$U=e^{iB}$ with $B=xu_1\otimes u_2$. Note  that
$\sigma_z\sigma_x\sigma_z^\dagger=-\sigma_x$ and
$\sigma_y\sigma_x\sigma_y^\dagger=-\sigma_x$. Then we can easily
check that $B=-ABA^\dagger$ for any $A\in {\cal U}$. Therefore, by
Fact 1, we have $U^\dagger=AUA^\dagger$.

 Next, we verify the necessity.  To do that, we first prove a
 fact  as follows.

 {\it Fact 2.}
 For Hermitian  operator $B$ on a $d$-dimensional Hilbert space, if
 $B=-ABA^\dagger$ holds for any $A\in\{\sigma_z\oplus I, \sigma_y\oplus
 I\}$, then $B$ necessarily has this form $B=c\sigma_x\oplus 0$ for some real number $c$.

 {\it Proof.} We assume  that $A=\sigma_z\oplus I$. Then  all the
 eigenvalues and eigenvectors of $A$ are  listed as follows
\begin{align*}
&A|0\rangle=|0\rangle,\\
&A|1\rangle=-|1\rangle,\\
&A|m\rangle=|m\rangle, ~~\text{for}~~ m=2,\dots, d-1.
\end{align*}
By this basis $\{|0\rangle,|1\rangle,\dots,|d-1\rangle\}$, $B$ can
be expressed in the outer product representation as follows
$$B=\sum_{ij}b_{ij}|i\rangle\langle j|.$$
Then by $B=-ABA^\dagger$, we can get that $B$ has the following
form
\begin{align}\begin{split}
B&=(b_{01}|0\rangle\langle1|+b_{10}|1\rangle\langle0|)\\
&+(\sum_{m=2}^{d-1}b_{1m}|1\rangle\langle
m|+\sum_{m=2}^{d-1}b_{m1}|m\rangle\langle 1|).\label{B11}
\end{split}
\end{align}
At the same time, we have $B=-A^{'}B{A^{'}}^\dagger$ for
$A^{'}=\sigma_y\oplus I$. Then substituting Eq. \eqref{B11} into
the right part of the equation $B=-A^{'}B{A^{'}}^\dagger$, we have
 \begin{align}\begin{split}
B&=(b_{01}|1\rangle\langle0|+b_{10}|0\rangle\langle1|)\\
&-i(\sum_{m=2}^{d-1}b_{1m}|0\rangle\langle
m|-\sum_{m=2}^{d-1}b_{m1}|m\rangle\langle
0|).\label{B12}\end{split}
 \end{align}
 Comparing Eqs.~\eqref{B11} and \eqref{B12}, we have that $b_{01}=b_{10}$
 and $b_{1m}=b_{m1}=0$ for $m=2,\dots, d-1$. Therefore, we have
 \begin{align}
 B=b_{10}(|1\rangle\langle0|+|0\rangle\langle1|)=c\sigma_x\oplus
 0.
 \end{align}
 Furthermore, from the Hermiticity of $B$,
  $c$ should be a real number. Hence, we end the proof of Fact 2. \qed

 Now we let $U=e^{iB}$ for some Hermitian
 operator $B$ on  $d\otimes d$, and suppose that
 $U^\dagger=AUA^\dagger$ holds for any $A\in{\cal U}$. Then from
 Fact 1, it follows that $B=-ABA^\dagger$ for any $A\in{\cal U}$. We should now prove
 that $B=xu_1\otimes u_2$.
 Denote $\{|0\rangle,\dots, |d-1\rangle\}$ the eigenvectors of $\sigma_z\oplus I$. Then $\{|i\rangle|j\rangle: i,j=0,\dots, d-1\}$ is a
 basis of the $d\otimes d$ Hilbert space.  Further, with this basis,  we can write $B$ in the following form
 \begin{align}
 B=\sum_{mn}|m\rangle\langle n|\otimes C_{mn}
 \end{align}
where
 \begin{align}
 C_{mn}=\sum_{ij}c^{mn}_{ij}|i\rangle\langle j|.
 \end{align}

 Assume that $A=(\sigma_z\oplus I)\otimes I$ in the equation
 $B=-ABA^\dagger$. Then we can get an equation similar to
 Eq.~\eqref{B11} as follows
 \begin{align}\begin{split}
 B&=(|0\rangle\langle1|\otimes C_{01}+|1\rangle\langle0|\otimes C_{10})\\
 &+(\sum_{m=2}^{d-1}|1\rangle\langle
m|\otimes C_{1m}+\sum_{m=2}^{d-1}|m\rangle\langle 1|\otimes
C_{m1}).\label{B21}\end{split}
 \end{align}
 At the same time, we have $B=-A^{'}B{A^{'}}^\dagger$ with $A^{'}=(\sigma_y\oplus I)\otimes
 I$. Then similar to Eq.~\eqref{B12}, we have
 \begin{align}\begin{split}
B&=(|1\rangle\langle0|\otimes C_{01}+|1\rangle\langle0|\otimes
C_{10})\\
&-i(\sum_{m=2}^{d-1}|0\rangle\langle m|\otimes
C_{1m}-\sum_{m=2}^{d-1}|m\rangle\langle 0|\otimes
C_{m1}).\label{B22}\end{split}
 \end{align}
 Comparing Eqs.~\eqref{B21} and \eqref{B22}, we have
 $C_{10}=C_{01}$ and $C_{1m}=C_{m1}=0$ for $m=2,\dots,d-1$.
 Therefore, we have
 \begin{align}
 B=(|1\rangle\langle0|+|0\rangle\langle1|)\otimes C_{01}=(\sigma_x\oplus 0)\otimes
 C_{01},
 \end{align}
 where $C_{01}$ should be a Hermitian operator on a $d$-dimensional Hilbert
 space, because of the Hermiticity of $B$.

 On the other hand, we have $B=-ABA^\dagger$ for
 $A\in\{I\otimes(\sigma_z\oplus I),I\otimes(\sigma_y\oplus I)\}$,
 which is equivalent to the following equation
 \begin{align}
 C_{01}=-A^{'}C_{01}{A^{'}}^\dagger ~\text{for}~ A^{'}\in\{\sigma_z\oplus
 I,\sigma_y\oplus I)\}.
 \end{align}
 Then, by Fact 2, we have $C_{01}=c\sigma_x\oplus0$.
 Therefore, we end the proof of Lemma 5 by proving  that $B=xu_1\otimes
 u_2$ for some real number $x$.\qed

 {\it Conclusion}---In this article, we have considered
the discrimination of two bipartite unitary operations by LOCC.
Specifically, we have shown that two bipartite unitary operations
acting on a $2$-qudit system can always be locally distinguished
by a sequential scheme without employing any entanglement,
improving the latest work in \cite{Duan2, Zhou}, as well. As a
result, we have obtained this statement: local entanglement is not
necessary for discriminating unitary operations acting on a
$2$-qudit system by LOCC, which is a stronger outcome than that in
\cite{Duan1}. Lastly, we have proposed some open problems for
further study: how to deal with the case of two subsystems having
different dimensions, how to extend our result to multipartite
unitary operations, and how to determine the optimal number of
runs $N$.

We thank Dr. X.F. Zhou for helpful discussions. This work is
supported  by the National Natural Science Foundation (Nos.
90303024, 60573006), and the Research Foundation for the Doctorial
Program of Higher School of Ministry of Education (No.
20050558015) of China.


\begin{thebibliography}{ABCD}
\bibitem {Hel} C. W. Helstrom, {\it Quantum Detection and Estimation
Theory} (Academic Press, New York, 1976).
\bibitem {Four}I.D. Ivanovic, Phys. Lett. A {\bf 123}, 257 (1987); D. Dieks, {\it ibid}.
{\bf 126}, 303 (1988); A. Peres, {\it ibid}. {\bf 128}, 19 (1988);
G. Jaeger and A. Shimony, {\it ibid}. {\bf 197}, 83 (1995); A.
Chefles, {\it ibid}. {\bf 239}, 339 (1998).

\bibitem{Bergou04}  J. Bergou, U. Herzog, and M. Hillery,
 {\it Quantum State Estimation}, Lecture Notes in Physics Vol. 649
(Springer, Berlin, 2004), p. 417; A. Chefles, {\it ibid}. p. 467.
\bibitem{Ben99} C. H. Bennett, D. P. DiVincenzo, C. A. Fuchs, T. Mor,
E. Rains, P. W. Shor, J. A. Smolin, and W. K. Wootters, Phys. Rev.
A {\bf 59}, 1070 (1999).
\bibitem{Walgate} J. Walgate, A.J. Short, L. Hardy, and V. Vedral, Phys. Rev.
Lett. {\bf 85}, 4972 (2000).
\bibitem {MFS2} M. F. Sacchi,  Phys. Rev. A {\bf 71}, 062340 (2005).
\bibitem{Chef07}  A. Chefles, A. Kitagawa, M. Takeoka, M. Sasaki, and J.
Twamley, quant-ph/0702245v3.
\bibitem {Childs} A. M. Childs, J. Preskill, and J. Renes, J. Mod. Opt. {\bf 47}, 155
(2000).
\bibitem {Acin} A. Ac\'{i}n, Phys. Rev. Lett. {\bf 87}, 177901 (2001).
\bibitem {Paris} G.M. D'Ariano, P. Lo Presti, and M.G.A. Paris,
Phys. Rev. Lett. {\bf  87}, 270404 (2001).
\bibitem {Duan1}R.Y. Duan, Y. Feng, and M. Ying,  Phys. Rev. Lett. {\bf 98}, 100503 (2007).
\bibitem {Duan2}R.Y. Duan, Y. Feng, and M. Ying,
quant-ph/0705.1424.
\bibitem{Zhou} X.F. Zhou, Y.S. Zhang, and G.C. Guo,
quant-ph/07051814.


\bibitem{Horn} R. A. Horn and C. R. Johnson, {\it Topics in Matrix Analysis}
(Combridge University Press, Cambridge, 1991).
 \bibitem{Bry} J.-L. Brylinski and R. Brylinski, {\it Mathematics of Quantum Computation}, edited by R. Brylinski and G. Chen (CRC Press,
Boca Raton, 2002) or see quant-ph/0108062.
 \bibitem{Meyer} D.A. Meyer, Phys. Rev. Lett.
{\bf 85}, 2014 (2000).
\bibitem{Rm} In the original protocol of \cite{Walgate}about the local distinguishability of two orthogonal bipartite
pure states, generally Bob's mesurement depends on the result of
Alice's measurement, and thus there needs some classical
comunication from Alice to Bob.  Besides, when the dimension of
Aice is not the powers of $2$, Alice should introudce an auxiliary
system whose dimension is the powers of $2$, and also she must
perform a collective unitary operation--- SWAP operation on the
systems holden by her. But for two orthogonal product states, a
perfect local discrimination can  be easily achived without the
classical commuication and auxiliary system involved.
\end{thebibliography}
\end{document}